\documentclass[prl,twocolumn,superscriptaddress,showpacs,nobalancelastpage]{revtex4}

\usepackage{graphicx}

\begin{document}

\title{Kondo universal scaling for a quantum dot coupled to superconducting leads}

\author{C. Buizert}
\affiliation{Dept. of Applied Physics, University of Tokyo, 7-3-1 Hongo, Bunkyo-ku, 113-8656, Japan}
\author{A. Oiwa}
\affiliation{Dept. of Applied Physics, University of Tokyo, 7-3-1 Hongo, Bunkyo-ku, 113-8656, Japan}
\author{K. Shibata}
\affiliation{IIS, University of Tokyo, 4-6-1 Komaba, Meguro-ku, Tokyo 153-8505, Japan}
\author{K. Hirakawa}
\affiliation{IIS, University of Tokyo, 4-6-1 Komaba, Meguro-ku, Tokyo 153-8505, Japan}
\author{S. Tarucha}
\email{tarucha@ap.t.u-tokyo.ac.jp}
\affiliation{Dept. of Applied Physics, University of Tokyo, 7-3-1 Hongo, Bunkyo-ku, 113-8656, Japan}
\affiliation{Quantum Spin Information Project, ICORP, Japan Science and Technology Agency, Atsugi-shi, Kanagawa 243-0198, Japan}

\date{\today}

\begin{abstract}
We study competition between the Kondo effect and superconductivity in a
single self-assembled InAs quantum dot contacted with Al lateral electrodes.
Due to Kondo enhancement of Andreev reflections the zero-bias anomaly
develops sidepeaks, separated by the superconducting gap energy $\Delta$. For ten
valleys of different Kondo temperature $T_K$ we tune the gap $\Delta$ with an external
magnetic field. We find that the zero-bias conductance in each case collapses
onto a single curve with $\Delta/k_{B}T_K$ as the only relevant energy scale, providing
experimental evidence for universal scaling in this system. 

\end{abstract}


\maketitle

The concept of universality evolved from the study of phase transitions in the field of statistical mechanics, where it was found that critical phenomena can be grouped in classes. All systems within the same universality class show identical behavior near the critical point for a certain set of relevant observables, despite the differences in microscopic details of the systems at hand \cite{Stanley}. Although formally not a phase transition, the Kondo effect has been shown to exhibit universality as well \cite{RalphScaling}. The Kondo effect arises from the anti-ferromagnetic interaction of localized magnetic impurities with conduction electrons. Below a characteristic temperature, the Kondo temperature $T_K$, a strongly correlated many-body state is formed that screens the impurity and greatly increases its scattering cross-section \cite{Hewson}. The last ten years have seen a renewed interest in this phenomenon with the possibility to use the unpaired electron spin on a quantum dot (QD) as the impurity, a setup that offers unprecedented control over experimental parameters \cite{GoldhNature, Cronenwett}. 

Kondo universality manifests itself as the governance of the Kondo energy scale $k_{B}T_K$ over all low-energy physics of the system. Other scales, such as the charging energy $U_C$ or lead coupling $\Gamma$, can be an order of magnitude larger, yet to describe the response to external parameters we require only $T_K$ and $G_0$, the zero temperature conductance. Under influence of a finite temperature, bias voltage or magnetic field ($X = k_{B}T$, $eV_{SD}$ or $g\mu_BB$) the normalized conductance $G(X)/G_0$ is reduced monotonically from unity following a curve that scales only with $X/k_{B}T_K$ \cite{Grobis}. Once appropriately scaled the behavior of any microscopic realization of the Kondo model can be collapsed onto this curve. The prime example of Kondo universality is its temperature dependence, where the empirical function \begin{equation}
  G(T)/G_0 = \left[ 1/(1 + (T/T_K^*)^2)  \right]^s
 \label{E_tempuniv}
 \end{equation}
 is widely used to fit experimental data. Here $T_K^* = T_K / \sqrt{2^{1/s} -1}$ , to conform with the definition of the Kondo temperature $G(T_K) = G_0 /2$, and $s = 0.22$ for spin-1/2 systems \cite{GoldhNRG}. Once expressed in the dimensionless parameters $G(T)/G_0$ and $T/T_K$ the behavior of e.g. semiconductor based dots \cite{GoldhNRG}, carbon nanotubes \cite{Nygard, Buitelaar} and single-molecule transistors \cite{Liang} could be mapped onto curve \ref{E_tempuniv}, showing the dominance of $T_K$ over the system's microscopic details.

Both superconductivity and the Kondo effect are many-body interactions with a strict inherent spin ordering. In a regular s-wave superconductor the electrons condense in singlet Cooper pairs, whereas the Kondo ground state consists of a dynamic many-body singlet formed between the localized dot electron and the itinerant lead electrons. Their competition is characterized by the relative strength $\Delta/k_{B}T_K$; for $\Delta/k_{B}T_K \ll 1$ the Kondo effect can survive in the presence of SC leads, in the opposite limit $\Delta/k_{B}T_K \gg 1$ Kondo is suppressed restoring Coulomb blockade (CB). The possibility of Andreev scattering at the dot-SC interface results in an even richer physics. Through Multiple Andreev Reflections (MAR) a Cooper pair of charge $2e$ can be transported across the dot \cite{MARSov}. In the Kondo regime a resonant channel opens up at the Fermi level, making the dot more transparent for the MAR processes. This allows the conductance to exceed the unitary limit of $2e^2/h$, provided that $T_K$ exceeds $\Delta$ \cite{AvishaiGolub}. This surprising prediction was confirmed experimentally by Buitelaar \textit{et al.} \cite{Buitelaar} for a carbon nanotube with Al electrodes. They observed a crossover between enhancement and suppression of the Kondo conductance at $k_{B}T_K \approx \Delta$, explained with a resistively-shunted junction model \cite{Choi}. We work in a different regime where MAR is strongly damped by repulsive Coulomb interactions, though we will see that first order AR is enhanced by the Kondo effect. In contrast to reference \cite{Buitelaar} the zero-bias conductance is suppressed over the entire range of Kondo temperatures by the gap $\Delta$ at the Fermi level.\\

\begin{figure}[!t]
\includegraphics[width=3.2in]{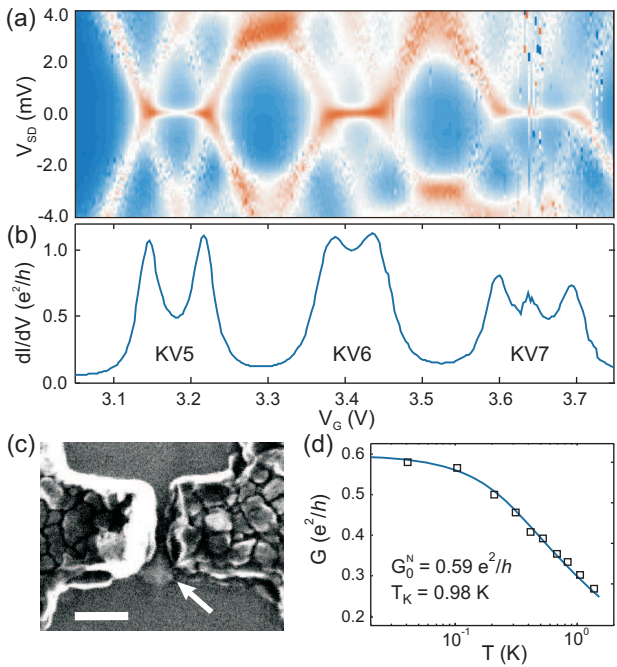}
\caption{(Color online) (a) Differential conductance $\textrm{d}I/\textrm{d}V_{SD}$ as a function of $V_G$ and $V_{SD}$, with the leads driven normal by a field $B = 100$ mT. $T = 35$ mK. (b) $\textrm{d}I/\textrm{d}V_{SD}$ at $V_{SD} = 0$ mV in units of $e^2/h$. We show KV5 to KV7 with a $T_K$ of 1.04, 1.65 and 1.18 K respectively. (c) Scanning Electron Micrograph of a similar device as used in these studies, with 100 nm scale bar. An InAs dot bridges the source and drain electrodes. (d) Temperature dependence of the conductance at KV5, with a best fit to scaling curve \ref{E_tempuniv}.}
\label{Fig1}
\end{figure}

We shall study transport through a single self-assembled InAs quantum dots contacted laterally via the nanogap method \cite{Hirak1} to Al SC electrodes. The dots are formed by MBE deposition of $\sim4$ ML InAs on a GaAs substrate, at this coverage many individual dots have coalesced into larger InAs islands with a diameter of $\sim80$ nm. The source and drain contacts, separated by a 25 nm gap, are defined by e-beam lithography. E-beam deposition of Ti (5 nm) and Al (100 nm) is preceded by a 5 s BHF etching step to de-oxidize the surface. Roughly 5\% of the gaps is bridged by a dot, yielding a working device (fig. \ref{Fig1}c). Each Al electrode is doubly contacted, allowing 4-terminal measurements. A degenerately doped layer 300 nm below the surface acts as an electrostatic backgate. Using a 4-terminal Al test strip deposited simultaneously we can characterize the Al leads to have a critical temperature $T_C = 0.9$ K and critical field $B_C = 95$ mT in perpendicular orientation. More details of the fabrication process will be given elsewhere \cite{Shibata}.

The measurements presented were done on a device with a RT junction resistance of 100 k$\Omega$. First we apply a 100 mT field to suppress superconductivity in the electrodes. Figure \ref{Fig1}a shows the charge stability diagram (Coulomb diamond) of the device, where we control the number of electrons on the dot with the voltage $V_G$ applied to the back gate. Figure \ref{Fig1}b shows the linear response at zero bias. We find a dot charging energy $U_C \sim 2.5$ meV and an electronic level spacing  $\delta E \sim 2.5$ meV. Counting the Coulomb oscillations we obtain an electron occupancy of $N=12$ for the leftmost valley at $V_G=3.0$ V. Around $V_G = 3.64$ V we reproducibly suffer from an electrostatic switch in the environment, which we attribute to the charging of a nearby InAs dot.

The stability diagram clearly exhibits the even-odd parity associated with spin-1/2 Kondo, showing an enhanced conductance ridge around zero bias for odd occupancy. For the various Kondo valleys the value of $T_K$ is different each time, as the electronic orbitals couple differently to the reservoirs. In addition to this variation, thermal cycling to room temperature also alters the dot parameters including $T_K$. Divided over three cooldowns of the device we were able to study 10 different values of $T_K$, ranging from 0.7 to 2 K. We shall refer to them as Kondo Valley (KV) 1-10.
For KV5 the zero-bias conductance is shown as a function of temperature in figure \ref{Fig1}d, the line shows a best fit to the empirical curve \ref{E_tempuniv}. The fitting gives $T_K = 0.98$ K, which is close to the value of 1.04 K found from half the width of the Kondo resonance out of equilibrium. From now on we shall use the latter method to determine $T_K$. The maximum normal state conductance $G_0^N$ was always smaller than $2e^2/h$, indicating an asymmetric lead coupling.

\begin{figure}[!t]
\includegraphics[width=3.2in]{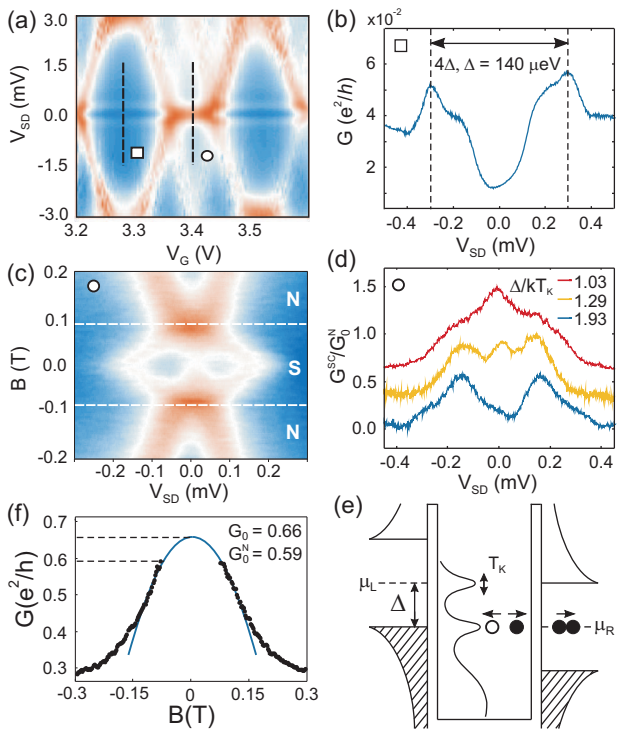}
\caption{(Color online) (a) $\textrm{d}I/\textrm{d}V_{SD}$ as a function of $V_G$ and $V_{SD}$, with the leads in the SC-state ($B = 0T$). (b) Scan of the gap structure in the CB regime, as indicated by the square in upper figure. Conductance is suppressed for $V_{SD}<2\Delta$. (c) Color-scale plot of $\textrm{d}I/\textrm{d}V_{SD}$, showing the field dependence of the Kondo resonance (KV2, $T_K = 1.23$ K). Horizontal lines denote the SC-N transition in the leads. (d) The Kondo resonance out of equilibrium for three different Kondo valleys (KV 10,2,4 from upper to lower) with the leads in the SC state. The curves are normalized by their N- state conductance $G_0^N$. Offset for clarity. (e) Schematic depiction of the Kondo enhanced Andreev Reflection, with a split Kondo resonance in the dot DOS. (f) Zero-bias Kondo conductance (KV2) as a function of magnetic field, with N-state leads.}
\label{Fig2}
\end{figure}

Sweeping back to zero magnetic field we bring the electrodes in the SC state. Figure \ref{Fig2}a shows the zero field Coulomb diamond. For even occupancy two parallel lines at source-drain (SD) bias $|V_{SD}| = 2\Delta/e$ mark the onset of direct quasiparticle tunneling. Their separation of $4\Delta$ allows us to determine $\Delta = 0.14$ meV, in good agreement with the critical temperature of the leads via the BCS relation  $\Delta(T=0) = 1.76k_BT_C$ \cite{Tinkham}. An accurate scan of the gap structure at even electron number is shown in Fig \ref{Fig2}b, the peaks at $|V_{SD}| = 2\Delta/e$ reflect the singularity in the density of states (DOS) of the leads. The MAR found in other QD devices connected to SC-leads \cite{BuitelaarMAR, PabloSCtransistor} is here absent due to strong on-site Coulomb interaction between the carriers ($U_C \sim 17\Delta$) \cite{Ralph95}. Only the first order Andreev reflection at $|V_{SD}| = \Delta/e$ is faintly visible as a shoulder to the $2\Delta$ feature.\\

We now turn to the behavior in the Kondo regime. The transition from the N- to SC-state with the magnetic field is shown in Fig \ref{Fig2}c. With the leads in the N-state the field splits the Kondo resonance; we observe a peak in the differential conductance when $V_{SD}$ equals the Zeeman energy \cite{GoldhNature}. A linear fit gives a g-factor $|g| = 6.6 \pm 0.3$. Below $B_C$ we surprisingly observe three resonances in the differential conductance. The two non-equilibrium ones occur at $|V_{SD}| = \Delta/e$, reflecting the SC DOS of the leads. Their position follows the field dependence of $\Delta$, being maximum at zero field. This conspicuous positioning corresponds to the first order Andreev reflection (AR), the physical process where an incoming electron scatters at the N-SC interface, injecting a Cooper pair in the SC-region and a backreflected hole in the N-region \cite{Tinkham}. Trough an interaction with the Kondo effect the AR is enhanced \cite{Vecino,Choi}; the mechanism is depicted schematically in Fig \ref{Fig2}e. In the middle of the valleys charge fluctuations are inhibited and the electrons traverse the dot through cotunneling processes. In the CB regime the lowest available state is $\sim (U_C + \delta E)/2$ away, but, by contrast, in the Kondo regime the many-body resonance gives rise to a finite DOS at the Fermi energy, enhancing the conductance. At finite bias the Kondo DOS splits with a maximum at each lead chemical potential, for $V_{SD} = \Delta/e$ this maximum aligns exactly with the band edge of the opposite reservoir. The electrons that contribute to the AR tunnel at this energy, and in this way the first order AR is strongly enhanced by the Kondo effect.

In Fig \ref{Fig2}d we compare the normalized conductance $G^{SC}/G_0^N$ as a function of $V_{SD}$ for 3 different valleys. Surprisingly the height $G^{SC}|_\Delta$ of the satellites appears to be insensitive to $\Delta/k_BT_K$, and was between 0.65 and 0.9$G^N_0$ for all 10 values of $T_K$ studied. A very recent work by Sand-Jespersen \cite{SandJesp} reports on similar Kondo enhanced AR in dots of small $T_K$. Why this process can persist where $\Delta \gg k_BT_K$ is uncertain. We speculate that the exact alingment with the band edge allows an out-of-equilibrium recovery of the Kondo state similar to the case of an applied magnetic field. Note that the direct tunneling onset at $|V_{SD}| = 2\Delta /e$ is not enhanced by Kondo, and its features are dwarfed by the much larger $\Delta$-peaks.
 
In contrast to the $\Delta$-satellites the zero-bias conductance is very sensitive to the relative strength $\Delta/k_{B}T_K$, or the ability of the Kondo effect to break up lead spin-pairing. For $\Delta \simeq k_{B}T_K$ (upper graph) the zero-bias conductance is close to its N-state value, whereas for higher $\Delta /k_{B}T_K$ the SC gap suppresses the conductance increasingly. Note that the resonance around $V_{SD} = 0$ cannot be interpreted as a supercurrent, as was done by Grove-Rasmussen \textit{et al.} \cite{Rasmus}. A low critical current $I_C \sim 100$ pA is to be expected as the Cooper pairs of charge $2e$ cannot tunnel directly due to the high $U_C$. The lead noise temperature of $\sim 100$ mK is roughly 10 times higher than the Josephson energy $E_J = \hbar I_C/2e$, destroying SC phase coherence across the junction. Indeed no dissipationless flow was detected in a current biased 4-terminal measurement.\\

\begin{figure}[!t]
\includegraphics[width=3.3in]{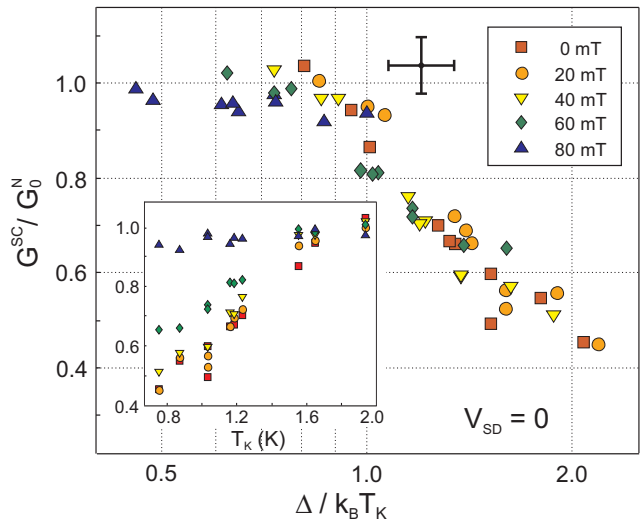}
\caption{(Color online) Semi-log plot of $G^{SC}/G_0^N$ for KV1-10 at various fields. Plotted against $\Delta/k_{B}T_K$ the data collapses on a single curve. The field dependent value of $\Delta$ was 135, 140, 120, 100 and 60 $\mu$eV for 0, 20, 40, 60 and 80 mT respectively. The measurement uncertainty is given by the bars. Inset shows the same data plotted against $T_K$ only.}
\label{Fig3}
\end{figure}

We analyze the 10 separate Kondo valleys with each a different $T_K$. On top of that we use the magnetic field to alter the SC gap, providing us a knob to experimentally tune $\Delta/k_{B}T_K$. By comparison, using the $V_G$ dependence of $T_K$ as a knob gives a much smaller range of variation. In Fig \ref{Fig3} we have plotted $G^{SC}/G_0^N$ at different fields as a function of $\Delta/k_{B}T_K$. When plotted in this fashion all data points collapse onto a single curve, proving that $\Delta/k_{B}T_K$ is the only relevant scaling parameter for the system. For comparison the inset shows the same data set as a function of just $T_K$. The scaling curves for other external parameters, such as temperature \cite{Pustilnik}, bias \cite{Rosch} or irradiation \cite{Kaminski} ($X = k_BT$, $eV_{SD}$ and $\hbar\omega$), are characterized by a logarithmic conductance fall-off in the weak coupling regime ($X \gg k_{B}T_K$) where a perturbative approach is valid. Although the data range is too limited to unambiguously claim any functional form, the observed fall-off seems to agree with the expected logarithmic dependence. Fitting the data with an empirical form such as function \ref{E_tempuniv} gives a poor correspondence. However, the $k_BT$ and $\Delta$ scaling curves need not be the same as the underlying mechanism is fundamentally different. Furthermore we believe the shape of the curve to be slightly distorted by the magnetic field, as discussed later on.
In contrast to the results of Buitelaar \cite{Buitelaar} we always find that the gap $\Delta$ suppresses the zero bias conductance, also in the regime $k_{B}T_K > \Delta$ . The main difference is the strong Coulomb repulsion in our device that damps the MAR, the latter being necessary to raise $G^{SC}$ above $G_0^N$. Unfortunately the field behavior is not discussed in reference \cite{Buitelaar}, making a direct comparison difficult.

To shed some light on the origin of the observed scaling we want to point out a similarity with the effect of other perturbations. Under equilibrium conditions the spin-flip cotunneling processes that create the Kondo singlet conserve energy. At finite $V_{SD}$ or $k_BT$ there is phase space available to inelastic processes, leading to decoherence of the many-body Kondo state \cite{Rosch,Kaminski}. In the case of SC leads a spin-flip with the leads requires the breaking of a Cooper pair, creating two excitations in the conduction band. This cost of $2\Delta$ requires energy exchange with the environment, leading to a suppressed conductance through decoherence of the Kondo singlet.

%Strong spin-orbit interaction, as in the case of InAs, can weaken the Kondo correlations by providing a fast spin relaxation mechanism. This might explain why in InAs nanowires the Kondo effect is often absent in regimes where it would otherwise be expected \cite{JordenSC}. However, in our devices the source-drain spacing of 25 nm is five times smaller than the spin-orbit length $\lambda_{SO}$ of $\sim 130$ nm determined for InAs nanowires \cite{Fasth_SO}, allowing the Kondo effect to survive.
Finally we want to address the non-negligible Zeeman splitting caused by the field that tunes $\Delta$. At 80 mT it equals $E_Z =$30 $\mu$eV, which should be compared to the Kondo energy $k_BT_K$ that ranges from 70 to 170 $\mu$eV. The suppression of the conductance by the splitting depends on both $|B|$ and $T_K$, and consequently we can expect the shape of the observed scaling function to be slightly distorted. Furthermore we used $G^N_0$ at 100 mT, rather than the `true' $G_0$, for normalization. In figure \ref{Fig2}f we have plotted the equilibrium conductance of KV2 as a function of magnetic field, with SC-leads regime omitted. In the absence of a proper scaling curve for the $B$-dependence we have fitted a parabola, and estimate that $G_0$ is 12\% higher than $G^N_0$. Because the field reduces both $G^{SC}$ and $G^N_0$ the 12\% represents an upper bound on the error. This is still smaller than the measurement uncertainty as indicated by the bars; we therefore believe the influence of the unwanted Zeeman splitting does not falsify our claim of Kondo universal scaling.\\

\begin{acknowledgments}
The authors would like to thank Mikio Eto, Christopher Yorke, Yshai Avishai, Wolfgang Belzig, and Leo Kouwenhoven for helpful discussions and/or comments, and Katsuharu Yoshida for technical support. We acknowledge financial support from SORST-JST, NanoQUINE and the Grant-in-Aid for Scientific Research A (No 40302799, No 18201027).
\end{acknowledgments}

%\bibliography{ens,ion,mx,qc,qi,qo,qm,txt}

\begin{references}

\bibitem{Stanley}
H.E. Stanley, Rev. Mod. Phys \textbf{71}, S358 (1999).

\bibitem{RalphScaling}
D.C. Ralph, A.W.W. Ludwig, Jan von Delft and R.A. Buhrman, Phys. Rev. Lett \textbf{72}, 1064 (1994).

\bibitem{Hewson}
A.C. Hewson, \textit{The Kondo Problem to Heavy Fermions}, Cambridge University Press, Cambridge (1993)

%\bibitem{GoldhNature}
%D. Goldhaber-Gordon, H. Shtrikman, D. Mahalu, D. Abusch-Magder, U. Meirav and M. A. Kastner, Nature \textbf{391}, 156 (1998)

%\bibitem{Cronenwett}
%S.M. Cronenwett, T.H. Oosterkamp and L.P. Kouwenhoven, A Tunable Kondo Effect in Quantum Dots, Science \textbf{281}, 540 (1998)
%
%\bibitem{vdWiel}
%W.G. van der Wiel, S. De Franceschi, T.Fujisawa, J.M. Elzerman, S. Tarucha and L.P. Kouwenhoven, Science \textbf{289}, 2105 (2000)
%
%\bibitem{Hirak1}
%M. Jung, K. Hirakawa, Y. Kawaguchi, S. Komiyama, S. Ishida and Y. Arakawa, Appl. Phys. Lett. \textbf{86}, 33106 (2005)

\bibitem{GoldhNature}
D. Goldhaber-Gordon \textit{et al.}, Nature \textbf{391}, 156 (1998)

\bibitem{Cronenwett}
S.M. Cronenwett \textit{et al.}, Science \textbf{281}, 540 (1998)

%\bibitem{vdWiel}
%W.G. van der Wiel \textit{et al.}, Science \textbf{289}, 2105 (2000)

\bibitem{Grobis}
M. Grobis \textit{et al.}, in \textit{Handbook of Magnetism and Advanced Magnetic Materials, Vol. 5} (Wiley,  2007)

\bibitem{GoldhNRG}
D. Goldhaber-Gordon \textit{et al.}, Phys. Rev. Lett. \textbf{81}, 5225 (1998)

%\bibitem{GoldhNRG}
%D. Goldhaber-Gordon, J. Göres, M.A. Kastner, H. Shtrikman, D. Mahalu and U. Meirav, Phys. Rev. Lett. \textbf{81}, 5225 (1998)
%

%\bibitem{CostiNRG}
%T.A. Costi, A.C. Hewson and V. Zlatic, J. Phys. Cond. Matter \textbf{6}, 2519 (1994)

\bibitem{Nygard}
J. Nygård, D.H. Cobden, and P.E. Lindelof, Nature \textbf{408}, 342 (2000)


\bibitem{Buitelaar}
M.R. Buitelaar, T. Nussbaumer and C. Sch\"{o}nenberger, Phys. Rev. Lett. \textbf{89}, 256801 (2002)

\bibitem{Liang}
W. Liang, M.P. Shores, M. Bockrath, J.R. Long, and H. Park, Nature \textbf{417},725 (2002)

\bibitem{MARSov}
A.F. Andreev, Sov. Phys. JETP \textbf{19}, 1228 (1964)

\bibitem{AvishaiGolub}
Y. Avishai, A. Golub and A.D. Zaikin, Phys. Rev. B \textbf{67}, 041301(R) (2003)

\bibitem{Choi}
M.-S. Choi, M. Lee, K. Kang, W. Belzig, Phys. Rev. B, \textbf{70}, 020502(R) (2004)

\bibitem{Hirak1}
M. Jung \textit{et al.}, Appl. Phys. Lett. \textbf{86}, 33106 (2005)

\bibitem{Shibata}
K. Shibata, C. Buizert, A.Oiwa, K. Hirakawa, S. Tarucha, \textit{unpublished}


\bibitem{Tinkham}
M. Tinkham, \textit{Introduction to superconductivity}, McGraw-Hill, Singapore, (1996)

\bibitem{BuitelaarMAR}
M.R. Buitelaar \textit{et al.} Phys. Rev. Lett. \textbf{91}, 57005 (2003)

\bibitem{PabloSCtransistor}
P. Jarillo-Herrero, J.A. van Dam and L.P. Kouwenhoven, Nature \textbf{439}, 953 (2006)

\bibitem{Ralph95}
D.C. Ralph, C.T. Black and M. Tinkham, Phys. Rev. Lett, \textbf{74}, 3241 (1995)

%\bibitem{Meir}
%Y. Meir, N.S. Wingreen and P.A. Lee, Phys. Rev. Lett. \textbf{70}, 2601 (1993)

\bibitem{Vecino}
E. Vecino \textit{et al.}, Sol. St. Comm. \textbf{131}, 625 (2004)

\bibitem{SandJesp}
T. Sand-Jespersen \textit{et al.}, arXiv:cond-mat/0703264 (2007)

\bibitem{Rasmus}
K. Grove-Rasmussen, H.I. Jørgensen, and P.E. Lindelof, arXiv:cond-mat/0601371v2 (2006)

%\bibitem{JordenSC}
%J.A. van Dam \textit{et al.}, Nature \textbf{442}, 667 (2006)

\bibitem{Pustilnik}
M. Pustilnik and L. Glazman, J. Phys.: Cond. Matter \textbf{16}, 513 (2004)

\bibitem{Rosch}
A. Rosch, J. Kroha and P. W\"{o}lfle, Phys. Rev. Lett. \textbf{87}, 156802 (2001)

\bibitem{Kaminski}
A. Kaminski, Y.V. Nazarov and L.I. Glazman, Phys. Rev. B \textbf{62}, 8154 (2000)

%\bibitem{Fasth_SO}
%C. Fasth, A. Fuhrer, L. Samuelson, V.N. Golovach, D. Loss, arXiv:cond-mat/0701161v1 (2007)


\end{references}

%\bibliographystyle{apsrev}

%\input{graininess.bblshort}

\end{document}